\documentclass[pra,aps,showpacs,onecolumn,twoside,superscriptaddress]{revtex4}

\usepackage{amsmath,amsfonts,amssymb,color,epsfig,graphics,graphicx,latexsym,revsymb,theorem,url,verbatim}

\newtheorem{definition}{Definition}
\newtheorem{proposition}[definition]{Proposition}
\newtheorem{lemma}[definition]{Lemma}

\newtheorem{theorem}[definition]{Theorem}
\newtheorem{corollary}[definition]{Corollary}
\newtheorem{conjecture}[definition]{Conjecture}

\newtheorem{remark}[definition]{Remark}
\newtheorem{example}[definition]{Example}
\newtheorem{problem}[definition]{Problem}

\def\squareforqed{\hbox{\rlap{$\sqcap$}$\sqcup$}}
\def\qed{\ifmmode\squareforqed\else{\unskip\nobreak\hfil
\penalty50\hskip1em\null\nobreak\hfil\squareforqed
\parfillskip=0pt\finalhyphendemerits=0\endgraf}\fi}
\def\endenv{\ifmmode\;\else{\unskip\nobreak\hfil
\penalty50\hskip1em\null\nobreak\hfil\;
\parfillskip=0pt\finalhyphendemerits=0\endgraf}\fi}
\newenvironment{proof}{\noindent \textbf{{Proof.~} }}{\qed}
\def\Dbar{\leavevmode\lower.6ex\hbox to 0pt
{\hskip-.23ex\accent"16\hss}D}
\makeatletter
\def\url@leostyle{%
  \@ifundefined{selectfont}{\def\UrlFont{\sf}}{\def\UrlFont{\small\ttfamily}}}
\makeatother
\def\bcj{\begin{conjecture}}
\def\ecj{\end{conjecture}}
\def\bcr{\begin{corollary}}
\def\ecr{\end{corollary}}
\def\bd{\begin{definition}}
\def\ed{\end{definition}}
\def\bea{\begin{eqnarray}}
\def\eea{\end{eqnarray}}
\def\bem{\begin{enumerate}}
\def\eem{\end{enumerate}}
\def\bex{\begin{example}}
\def\eex{\end{example}}
\def\bim{\begin{itemize}}
\def\eim{\end{itemize}}
\def\bl{\begin{lemma}}
\def\el{\end{lemma}}
\def\bpf{\begin{proof}}
\def\epf{\end{proof}}
\def\bpp{\begin{proposition}}
\def\epp{\end{proposition}}
\def\br{\begin{remark}}
\def\er{\end{remark}}
\def\bt{\begin{theorem}}
\def\et{\end{theorem}}

\newcommand{\nc}{\newcommand}

\def\a{\alpha}
\def\b{\beta}

\def\l{\lambda}

\def\r{\rho}
\def\s{\sigma}

\def\ph{\varphi}

\def\ps{\psi}

\def\G{\Gamma}

\nc{\bbC}{{\mathbb{C}}}

\nc{\cA}{{\cal A}} \nc{\cB}{{\cal B}} \nc{\cC}{{\cal C}}
\nc{\cD}{{\cal D}} \nc{\cE}{{\cal E}} \nc{\cF}{{\cal F}}
\nc{\cG}{{\cal G}} \nc{\cH}{{\cal H}} \nc{\cI}{{\cal I}}
\nc{\cJ}{{\cal J}} \nc{\cK}{{\cal K}} \nc{\cL}{{\cal L}}
\nc{\cM}{{\cal M}} \nc{\cN}{{\cal N}} \nc{\cO}{{\cal O}}
\nc{\cP}{{\cal P}} \nc{\cR}{{\cal R}} \nc{\cS}{{\cal S}}
\nc{\cT}{{\cal T}} \nc{\cU}{{\cal U}} \nc{\cV}{{\cal V}}
\nc{\cW}{{\cal W}} \nc{\cX}{{\cal X}} \nc{\cZ}{{\cal Z}}

\nc{\hA}{{\hat{A}}} \nc{\hB}{{\hat{B}}} \nc{\hC}{{\hat{C}}}
\nc{\hD}{{\hat{D}}} \nc{\hE}{{\hat{E}}} \nc{\hF}{{\hat{F}}}
\nc{\hG}{{\hat{G}}} \nc{\hH}{{\hat{H}}} \nc{\hI}{{\hat{I}}}
\nc{\hJ}{{\hat{J}}} \nc{\hK}{{\hat{K}}} \nc{\hL}{{\hat{L}}}
\nc{\hM}{{\hat{M}}} \nc{\hN}{{\hat{N}}} \nc{\hO}{{\hat{O}}}
\nc{\hP}{{\hat{P}}} \nc{\hR}{{\hat{R}}} \nc{\hS}{{\hat{S}}}
\nc{\hT}{{\hat{T}}} \nc{\hU}{{\hat{U}}} \nc{\hV}{{\hat{V}}}
\nc{\hW}{{\hat{W}}} \nc{\hX}{{\hat{X}}} \nc{\hZ}{{\hat{Z}}}

\def\diag{\mathop{\rm diag}}
\def\dim{\mathop{\rm Dim}}

\def\max{\mathop{\rm max}}

\def\rank{\mathop{\rm rank}}

\def\tr{\mathop{\rm Tr}}

\def\GL{{\mbox{\rm GL}}}

\def\Un{{\mbox{\rm U}}}

\def\ox{\otimes}

\def\sue{\subseteq}

\newcommand{\bra}[1]{\langle#1|}
\newcommand{\ket}[1]{|#1\rangle}
\newcommand{\proj}[1]{| #1\rangle\!\langle #1 |}

\newcommand{\braket}[2]{\langle#1|#2\rangle}

\newcommand{\jmp}{J. Math. Phys.}
\newcommand{\jpa}{J. Phys. A}

\def\bR{{\mbox{\bf R}}}

\def\bC{{\mbox{\bf C}}}

\begin{document}
\title[Dimensions, lengths and separability]
{Dimensions, lengths and separability in finite-dimensional
quantum systems}

\author{Lin Chen}
\affiliation{Department of Pure Mathematics and Institute for
Quantum Computing, University of Waterloo, Waterloo, Ontario, N2L
3G1, Canada} \affiliation{Centre for Quantum Technologies, National
University of Singapore, 3 Science Drive 2, Singapore 117542}
\email{cqtcl@nus.edu.sg (Corresponding~Author)}

\def\Dbar{\leavevmode\lower.6ex\hbox to 0pt
{\hskip-.23ex\accent"16\hss}D}
\author {{ Dragomir {\v{Z} \Dbar}okovi{\'c}}}

\affiliation{Department of Pure Mathematics and Institute for
Quantum Computing, University of Waterloo, Waterloo, Ontario, N2L
3G1, Canada} \email{djokovic@uwaterloo.ca}

\begin{abstract}
Many important sets of normalized states in a multipartite
quantum system of finite dimension $d$, such as the set $\cS$
of all separable states, are real semialgebraic sets.
We compute dimensions of many such sets in several
low-dimensional systems. By using dimension arguments,
we show that there exist separable states which are not
convex combinations of $d$ or less pure product states.
For instance, such states exist in bipartite $M\ox N$ systems
when $(M-2)(N-2)>1$. This solves an open problem proposed
by DiVincenzo, Terhal and Thapliyal about 12 years ago.
We prove that there exist a separable state $\r$ and a pure product state, whose mixture has smaller length than that of
$\r$.
We show that any real $\r\in\cS$, which is invariant under all  partial transpose operations, is a convex sum of real pure
product states. In the case of the $2\ox N$ system, the number
$r$ of product states can be taken to be $r=\rank\r$.
We also show that the general multipartite separability problem
can be reduced to the case of real states. Regarding the
separability problem, we propose two conjectures describing
$\cS$ as a semialgebraic set, which may eventually lead to an
analytic solution in some low-dimensional systems such as
$2\ox4$, $3\ox3$ and $2\ox2\ox2$.
\end{abstract}

\date{ \today }




\maketitle



\section{\label{sec:introduction} Introduction}

Entanglement reveals a fundamental difference between the quantum and classical world, which may be detected e.g. by Bell inequallities \cite{bell64}.
It plays an essential role in quantum information processing,
such as quantum teleportation, computing and cryptography.
It is a hard problem to decide whether a given quantum state is entangled. We shall propose a new method of atacking this
problem in low-dimensional quantum systems,
see Conjectures \ref{Hip-1} and \ref{Hip-2}.

We consider a finite-dimensional multipartite quantum system
described by the complex Hilbert space
$\cH=\cH_1\ox\cdots\ox\cH_n$.
We set $d=d_1d_2\cdots d_n$ where $d_i=\dim\cH_i$.
For brevity, we may refer to this system also as an
$d_1\ox\cdots\ox d_n$.
A {\em (normalized) state} is a positive semidefinite linear
operator $\r:\cH\to\cH$ with $\tr\r=1$.
If the condition $\tr\r=1$ is omitted and $\r\ne0$, we shall say
that $\r$ is a {\em non-normalized state}.
For convenience, we prefer to work with non-normalized states.
We shall mention explicitly when we require the states to be
normalized.
A {\em pure product state} is the tensor product
$\proj{\psi_1}\ox\cdots\ox\proj{\psi_n}$, where
$\ket{\psi_i}\in\cH_i$ are nonzero vectors.
A state is {\em separable} if it is a finite sum of pure product
states.
A state is {\em entangled} if it is not separable.
There exists a necessary and sufficient condition for separability \cite{hhh96} which shows that the separability problem is equivalent to another hard problem about positive linear maps.
It has been shown that, in general, the problem of deciding
whether a state is entangled is NP hard \cite{gurvits03}.

Let $\cD$ be the set of all normalized states on $\cH$, and $\cS$
the subset of all normalized separable states. Let $H$ denote the
real vector space consisting of all Hermitian operators on $\cH$.
Denote by $H'$ the affine subspace of $H$ defined by the equation
$\tr\r=1$. Our starting point is the well known fact that $\cD$ is a
real semialgebraic subset of $H'$, see Section
\ref{sec:SemiAlgSets}. From now on we shall say ``semialgebraic"
instead of ``real semialgebraic". For the definition and examples of real algebraic and semialgebraic subsets of $\bR^n$ see e.g.
\cite[Section 2.1]{Boch98}. Let us just mention that
$H_r:=\{\r\in H:\rank\r\le r\}$ is a real algebraic set, and so
is the set $H'_r=H'\cap H_r$. Since finite intersections of
semialgebraic sets are semialgebraic, it follows that the sets
$\cD_r=\cD\cap H_r$ are semialgebraic.
We define the {\em length}, $L(\r)$, of any separable state $\r$
to be the least integer $r$ such that $\r$ can be written as a sum of $r$ pure product states. In Proposition
\ref{prop:SemiAlgSet} we prove that the sets
$\cS'_r:=\{\r\in\cS:L(\r)\le r\}$ are semialgebraic. From this
result we deduce (see Corollary \ref{cor:SemiAlgSet}) that $\cS$
is a semialgebraic set, and so are the sets $\cS_r=\cS\cap H_r$.

Any closed semialgebraic set is a finite union of basic closed
semialgebraic sets \cite[Theorem 2.7.2]{Boch98}. Thus $\cS$ is a
finite union of pieces, where each piece is described by a finite
number of polynomial inequalities of the type $f(\r)\ge0$ with
$f:H'\to\bR$ a polynomial function. We conjecture that in the
case of $\cS$ there is just one of these pieces, i.e., that $\cS$
is itself a basic closed semialgebraic set, see Definition
\ref{def:Basic} and Conjecture \ref{Hip-1}. A stronger version of
this conjecture is stated as Conjecture \ref{Hip-2}.
It asserts that the polynomial functions $f$ in the inequalities
$f(\r)\ge0$ defining $\cS$ can be chosen to be invariant under
the action of the unitary groups of the $\cH_i$.
If the latter conjecture is true, then it is likely that
the separability problem will be solved analytically in some
low-dimensional cases where the Peres-Horodecki criterion fails.

We denote by $\G_i$ the partial transposition operator on the
space $\cH_i$, $i=1,\ldots,n$, computed in some fixed o.n.
basis of that space. Thus, if
$\r=\r_1\ox\cdots\ox\r_i\ox\cdots\ox\r_n$ then
$\r^{\G_i}=\r_1\ox\cdots\ox\r_i^T\ox\cdots\ox\r_n$.
We denote by $G$ the group generated by the pairwise commuting involutory operators $\G_i$.
The elements of $G$ are the products $\G_S=\prod_{i\in S}\G_i$
where $S$ runs through all subsets of $\{1,2,\ldots,n\}$.
We say that a state $\r$ on $\cH$ {\em has positive partial
transposes} (or that it is PPT) if $\r^{\G_S}\ge0$ for all
subsets $S$. If a state $\r$ is not PPT, we shall say that it
is NPT.
It is immediate from the definition of separable states that
every separable state $\r$ is PPT.

We consider often the bipartite case $(n=2)$. In that case
we set $\cH_A=\cH_1$, $\cH_B=\cH_2$, $M=d_1$ and $N=d_2$.
We also set $\G=\G_1$.
We say that a bipartite state $\r$ is a $k\times l$ {\em state}
if its local ranks are $k$ and $l$, i.e., $\rank\r_A=k$ and $\rank\r_B=l$ where $\r_A=\tr_B(\r)$ and $\r_B=\tr_A(\r)$.
Let $\r$ be a $k\times l$ state.
If $k=1$ or $l=1$ then $\r$ is clearly separable (and PPT).
More generally, it is well known that $\r$ is separable
if it is PPT and $(k-1)(l-1)\le2$ \cite{peres96,hhh96}.

Due to the fact that the sets $\cD_r$, $\cS_r$ and $\cS'_r$ are
semialgebraic, they have well defined dimensions. We have
computed the dimension of $\cD_r$ for all $r=1,\ldots,d$, see Theorem \ref{thm:DimD_r}.
We also found tight upper bounds for $\dim\cS'_r$.
When these bounds do not exceed $d^2-1$, the dimension of the
ambient affine space $H'$, they are saturated in most cases.
For several low-dimensional quantum systems we have determined
the $\dim\cS'_r$ exactly, see Table \ref{tab:dimenzije}.

As a simple consequence of these bounds, we prove that if
$(M-2)(N-2)>1$ then there exist separable states on $M\ox N$
of length exceeding $MN$.
(E.g. in $3\ox4$, there exist separable states of lengths $13$
and $14$.)
This question was raised about 12 years ago \cite{dtt00} and
remained open until now.
In Conjecture \ref{conj:Granica}, we propose a new candidate
for the maximum length of separable states.
(A counter-example to this conjecture was recently found
by K-C. Ha and S-H. Kye \cite{hk12}.)
Apparently, it is much harder to compute $\dim\cS_r$.
We were able to do that in $2\ox2$ and $2\ox3$ systems, see
Proposition \ref{prop:dimS_r}.

We investigate the set $\cS^{\rm re}$ of separable density
matrices $\r\in\cS$ all of whose entries are real. We denote by
$\cS^G$ the set of $G$-invariant matrices $\r\in\cS$.
If a Hermitian operator $\r$ is $G$-invariant, then
$\r^T=\r$ and so $\r^*=\r$. i.e., $\r$ is real.
Thus, we have $\cS^G\sue\cS^{\rm re}$.
In Proposition \ref{pp:GamaInv} we show that the set $\cS^G$
consists of all states $\r$ which admit a representation as
a convex linear combination of normalized real pure product
states.
We denote by $L^G(\r)$ the minimal number of real pure product
states in such a convex linear combination.
In Theorem \ref{thm:2oxN,Ginvariant} we prove that, in the
$2\ox N$ system, we always have $L^G(\r)=\rank\r$.
The dimensions of the sets $\cS^{\rm re}$ and $\cS^G$ are
computed in Proposition \ref{pp:Dimenzije}.
In analogy to the sets $\cS'_r$, we introduce the sets
$\cS^G_r=\{\r\in\cS:L^G(\r)\le r\}$.
For bipartite systems of dimension $d\le16$, we give in
Table \ref{tab:SepDim} the lower bounds for dimensions of the
sets $\cS^G_r$ for $r=1,\ldots,d$. In most cases it is shown
that the bound is actually equal to $\dim\cS^G_r$.
We consider the following three separability problems
$(S_1)$, $(S_2)$ and $(S_3)$: for a given state $\r$, decide
whether $\r$ belongs to $\cS$, $\cS^{\rm re}$ or $\cS^G$,
respectively. Since $\cS\supseteq\cS^{\rm re}\supseteq\cS^G$,
$(S_2)$ is a special case of $(S_1)$, and $(S_3)$ a
special case of $(S_2)$.
We prove in Proposition \ref{pp:reduction-2} that $(S_1)$,
$(S_2)$ and $(S_3)$ are equivalent to each other.

The content of the paper is as follows. In section
\ref{sec:SemiAlgSets} we prove that the sets $\cD_r$, $\cS_r$ and
$\cS'_r$ are semialgebraic and propose the two separability
conjectures. In section \ref{sec:dimensions} we compute
$\dim\cD_r$, find a tight upper bound for $\dim\cS'_r$, and in some cases prove that they are equal.
In section \ref{RealSepSta} we characterize the set $\cS^G$
and compute its dimension.
Finally, in section \ref{sec:conclusion} we highlite some of
our results and discuss the prospects of solving analytically
the separability problem in some quantum systems of low
dimension.

\section{\label{sec:SemiAlgSets}
Some semialgebraic sets and separability conjectures}

The sets $\cD,\cS,\cD_r,\cS_r$ and $\cS'_r$ are compact subsets
of $H'$, and the first two are also convex.
It is not hard to prove that all of them
are also semialgebraic subsets of $H'$.

Let us recall and show that $\cD$ is a semialgebraic subset of
$H'$. Indeed, if $f(t)=\sum_{i=0}^d (-1)^i c_i t^{d-i}$,
$(c_0=1)$, is the characteristic polynomial of $\r\in H'$,
then $\r\in\cD$ if and only if each $c_i\ge0$.
Recall that $c_i$ is the sum of all principal minors of order
$i$ of $\r$, and so $c_i$ is a polynomial function on $H'$.
Thus, $\cD$ is semialgebraic.
In fact this shows that $\cD$ is a basic closed semialgebraic
set according to the following definition.

 \bd \label{def:Basic} (see \cite[Definition 2.7.1]{Boch98})
A subset $X$ of a Euclidean space $E$ is a {\em basic
closed semialgebraic set} if there exist finitely many real
polynomial functions $f_i:E\to\bR$, $i=1,\ldots,k$, such that
$X=\{x\in E:f_i(x)\ge0,~i=1,\ldots,k\}$.
 \ed
(We shall use this definition when $E$ is either $H$ or $H'$.)

Since  $H_r$ is a real algebraic set and $\cD_r=\cD\cap H_r$, it follows that $\cD_r$ is semialgebraic. We shall now prove that this is also true for the sets $\cS'_r$ and $\cS_r$.

\bpp \label{prop:SemiAlgSet} Each set $\cS'_r$ is a semialgebraic subset of $H'$. \epp
\bpf Let $X_i$ be the unit sphere in $\cH_i$ and let the subset
$\Lambda\subset\bR^r$ be defined by the equality
$\l_1+\cdots+\l_r=1$ and the inequalities $0\le\l_i\le1$ for
$i\in\{1,\ldots,r\}$.
The spheres $X_i$ are real algebraic sets and so is their
product $X=X_1\times\cdots\times X_n$. As the set $\Lambda$ is
semialgebraic, the product
 \bea \label{ProstorZ}
Z=\Lambda\times X^r
 \eea
of $\Lambda$ and $r$ copies of $X$ is also semialgebraic.
We shall now define a map $f:Z\to \cS'_r$.
We shall write an arbitrary point $z\in Z$ as
$z=(\l_1,\ldots,\l_r,x_1,\ldots,x_r)$, where each
$x_i=(x_{i1},\ldots,x_{in})\in X$.
The function $f$ is defined by setting
$f(z)=\sum_{i=1}^r \l_i\proj{x_i}\in \cS'_r$, where
$\proj{x_i}=\proj{x_{i1}}\ox\cdots\ox\proj{x_{in}}$
is a pure product state.
The assertion follows from \cite[Theorem 2.8.8]{Boch98} since
$f$ is a real polynomial map and $f(Z)=\cS'_r$.
 \epf

 \bcr \label{cor:SemiAlgSet}
The set $\cS$ of all normalized separable states on $\cH$ is
semialgebraic, and so are its subsets $\cS_r$.
 \ecr
 \bpf
Since $\cS_r=\cS\cap \cD_r$, the second assertion follows from
the first. We shall give two proofs for the first assertion.

First proof: By a result of P. Horodecki
\cite[Theorem 1]{horodecki97}, which easily extends to the
multipartite case, we have $\cS=\cS'_r$ for $r=d^2$ and the
assertion follows from Proposition \ref{prop:SemiAlgSet}.

Second proof: By Proposition \ref{prop:SemiAlgSet}, $\cS_1$ is
semialgebraic. We can now apply the known fact that the convex
hull of a semialgebraic set is also semialgebraic. As $\cS$ is
the convex hull of $\cS_1$, we deduce that $\cS$ is
semialgebraic.
 \epf

Our first conjecture says that $\cS$ is a very nice
semialgebraic subsets of $H'$.

 \bcj \label{Hip-1}
The set $\cS$ is a basic closed semialgebraic subset of $H'$.
 \ecj

Our second conjecture is a stronger version of this one.
It asserts that the polynomial functions, which
occur in the representation of $\cS$ as a basic closed
semialgebraic set, can be chosen in a very special way.
In this place, it is convenient to work with the non-normalized states.
If $X\subseteq H'$ is a nonempty subset, we define the
{\em cone} over $X$ to be the subset
$KX=\{t\r:t\ge0,~\r\in X\}$ of $H$. Note that the vertex of
$KX$, namely the origin of $H$, belongs to $KX$. Consequently, if
$X$ is closed and compact, then $KX$ is closed. Note also that
if $X$ is convex or semialgebraic then $KX$ has the same
property, and conversely. We are in particular
interested in the cone $K\cS$ consisting of all non-normalized separable states (plus the origin).

The direct product of the general linear groups
$\GL:=\prod_{i=1}^n\GL_{d_i}(\bC)$ acts naturally on $\cH$ via local invertible transformations, and also acts on $H$. Explicitly, if $V\in\GL$ and $\r\in H$ then the latter action is
given by $(V,\r)\to V\r V^\dag$.
The {\em local unitary group}, i.e., the subgroup
$\Un=\prod_{i=1}^n\Un(d_i)$ of $G$, also acts on the same
spaces.
We say that a polynomial function $f:H\to\bR$ is
{\em invariant}, if $f(V\r V^\dag)=f(\r)$ for all $V\in\Un$
and $\r\in H$.
The Conjecture \ref{Hip-1} implies that the cone $K\cS$ is
a basic closed semialgebraic subset of $H$. The following
conjecture is much stronger.

 \bcj \label{Hip-2}
There exist finitely many homogeneous invariant polynomial
functions $f_i:H\to\bR$, $i=1,\ldots,k$, such that
$K\cS=\{\r\in H:f_i(\r)\ge0,~i=1,\ldots,k\}$.
 \ecj

Note that in the bipartite case both conjectures are
true if $(M-1)(N-1)\le2$. In spite of the fact that this has
been known for long time (albeit not stated in this way), all
other bipartite cases still remain unsolved.
If Conjecture \ref{Hip-2} is true, then it should be
possible to find analytic criteria of separability in some
additional low-dimensional cases, say for $2\ox4$, $3\ox3$ and
$2\ox2\ox2$ quantum systems.
To realize this objective, it is first of all necessary to find
a practical method for computing the homogeneous invariants
of small degree. Since $\Un$ is compact, the algebra of
polynomial invariants is finitely generated and the generators
can be chosen to be homogeneous.
We pose the following problem which would provide a simple
method for computing all homogeneous invariants.

\begin{problem}
Find a minimal set of homogeneous generators for the algebra
of invariant polynomials $f:H\to\bR$ for the quantum systems
$2\ox4$, $3\ox3$ and $2\ox2\ox2$.
\end{problem}

\section{\label{sec:dimensions}
Dimension computations for some quantum systems}

Since all of the sets $\cD_r$, $\cS'_r$ and $\cS_r$ are semialgebraic,
they have a well defined notion of dimension.
We shall compute some of these dimensions in certain cases.
We start with the sets $\cD_r$ in which case we can ignore
the tensor product structure of $\cH$.

 \bt \label{thm:DimD_r}
We have $\dim \cD_r=r(2d-r)-1$ for $r=1,2,\ldots,d$.
 \et
 \bpf
Any $\r\in \cD_r$ has a spectral decomposition
\bea \label{SpekRaz}
\r=\sum_{i=1}^r \l_i\proj{\psi_i}, \quad \sum\l_i=1, \quad
\l_i\ge0,\quad \|\psi_i\|=1.
\eea
We fix an o.n. basis $\ket{\a_k}$, $k=1,\ldots,d$ of $\cH$
and denote by $\Un(d)$ the global unitary group of $\cH$
with respect to this basis.
This group acts naturally on $\cH$ as well as on $H$ and $H'$
and its subset $\cD$. For $U\in\Un(d)$ and $\r\in H$, we shall
use the notation $U\cdot\r=U\r U^\dag$.
Denote by $X$ the set of all states $\r$ given by
(\ref{SpekRaz}) with $\ket{\psi_i}=\ket{\a_i}$, $i=1,\ldots,r$.
Clearly, we have $\cD_r=\Un(d)\cdot X$.
Let $Y\subset X$ consist of all $\r$ which also satisfy the
inequalities $\l_1>\l_2>\cdots>\l_r$.
The stabilizer in $\Un(d)$ of any state $\r\in Y$ is the
subgroup $\Un(1)^r\oplus\Un(d-r)$ of dimension $r+(d-r)^2$.
Hence the dimension of the orbit $\Un(d)\cdot\r$ is equal to
$r(2d-r)-r$. Since $\dim Y=r-1$, we infer that
$\dim \Un(d)\cdot Y=r(2d-r)-1$. As the closure of
$\Un(d)\cdot Y$ contains $X$, it must be equal to
$\Un(d)\cdot X=\cD_r$. Since $\Un(d)\cdot Y$ is a semialgebraic set, it follows that also $\dim \cD_r=r(2d-r)-1$.
 \epf

Note that for $r=d$ we recover the well known fact that
$\dim \cD=\dim H'=d^2-1$.

 \bt \label{thm:Granica}
For all positive integers $r$ we have
 \bea
\dim \cS'_r\le r(1+2\sum_i (d_i-1))-1.
 \eea
Consequently, there exist separable states of length
 \bea \label{frm:Granica}
l:=\bigg\lceil \frac{d^2}{1+2\sum_i (d_i-1)} \bigg\rceil,
 \eea
where $\lceil t \rceil$ denotes the smallest integer $\ge t$.
 \et
 \bpf
The space $Z$ (see Eq. (\ref{ProstorZ})) and
the map $f:Z\to\cS'_r$ provide a parametrization of the
set $\cS'_r$. This parametrization is redundant in the sense that
the overall phases of the unit vectors $\ket{\a_i}$ and
$\ket{\b_i}$ are irrelevant.
To obtain a more economical parametrization of $\cS'_r$, we
replace each sphere $X_i$ by its $(2d_i-2)$-dimensional
subsphere, $X_i^0$, consisting of all unit vectors whose first
component is real.
Let $X^0=X_1^0\times\cdots\times X_n^0$ and
$Z_0=\Lambda\times(X^0)^r$, and let
$f_0:Z_0\to \cS'_r$ be the restriction of the map $f$ used in the
proof of Proposition \ref{prop:SemiAlgSet}.
Since we still have $f_0(Z_0)=\cS'_r$,
it follows that $\dim\cS'_r\le\dim Z_0=r(1+2\sum_i (d_i-1))-1$.

If $r=l-1$ then $r(1+2\sum_i (d_i-1))<d^2$, and so
$\dim \cS'_r<d^2-1=\dim \cS$.
Thus $\cS'_r$ is a proper subset of $\cS$, and so there exist
$\r\in \cS$ with $L(\r)\ge l$.
 \epf

About 12 years ago the authors of \cite{dtt00} raised the question
whether the separable states on $M\otimes N$ satisfy the inequality
$L(\r) \le MN$. It follows from the theorem that the answer to this
question is negative. Indeed, if $(M-2)(N-2)>1$ then $\dim\cS=M^2
N^2-1>MN(2M+2N-3)-1\ge\dim\cS'_{MN}$ and so there must exist
separable states on $\cH$ of length bigger than $MN$.

It is now easy to answer the analogous question in the multipartite
case.
 \bcr
 \label{cr:L<=d}
Assume that $d_1\ge d_2\ge\cdots\ge d_n\ge2$ and $n\ge2$.
If $L(\r)\le d$ for all separable states $\r$ on $\cH$,
then $n=2$ and either $d_1=d_2=3$ or $d_2=2$.
 \ecr
 \bpf
By the theorem we have $l\le d$ and Eq. (\ref{frm:Granica})
implies that $d\le 1+2\sum_i (d_i-1)$.
Suppose that $n>2$. Then the function
$f(d_1,\cdots,d_n)=\prod_i d_i-1-2\sum_i (d_i-1)$
is strictly increasing as a function of a single variable $d_i$.
As $f(2,\ldots,2)=2^n-2n-1>0$, we have a contradiction.
We conclude that $n=2$, and the corollary follows easily.
 \epf

On the bipartite system $M\ox N$, if $M=1$ or $N=1$, all states are separable and so $\cS'_r=\cS_r=\cD_r$ and
$\dim\cD_r=r(2N-r)-1$ by Theorem \ref{thm:DimD_r}.
In Table \ref{tab:dimenzije}, we give the dimensions of the
sets $\cS'_r$ for several small values of $n$ and the $d_i$.
As $\cS'_1\sue\cS'_2\sue\cS'_3\sue\cdots$, we have
$\dim\cS'_1\le\dim\cS'_2\le\dim\cS'_3\le\cdots$.

Let us sketch the proof of the results stated in
Table \ref{tab:dimenzije}.
By Theorem \ref{thm:Granica} we have
$\dim K\cS'_r\le(2M+2N-3)r$. If $r$ is in the initial range
(as secified in Table \ref{tab:dimenzije}), we claim that
the equality holds.
Define the map $\ph:\cH_A\times\cH_B\to H$ by
$\ph(a,b)=\proj{a}\ox\proj{b}$ and observe that the rank of
its Jacobian matrix does not exceed $2M+2N-3$.
Next define the map $\Phi_r:(\cH_A\times\cH_B)^r\to H$ by setting
 \bea \label{eq:Phi}
\Phi_r(a_1,b_1,\ldots,a_r,b_r)=\sum_{i=1}^r \ph(a_i,b_i).
 \eea
The image of $\Phi_r$ is exactly the cone $K\cS'_r$.
Since $\Phi_r$ is a smooth map, the dimension of $K\cS'_r$
must be greater than or equal to the maximum rank of the
Jacobian matrix, $J[\Phi_r]$, of the map $\Phi_r$.
Hence, in order to prove the claim it suffices to find a point
$p_r:=(a_1,b_1,\ldots,a_r,b_r)\in(\cH_A\times\cH_B)^r$ such that
the rank of  $J[\Phi_r]$ at $p_r$ is equal to
the upper bound $(2M+2N-3)r$. Note that for $r>1$ we have
$J[\Phi_r]=J[\Phi_{r-1}\circ\pi]+J[\ph\circ\pi']$, where
$\pi:(\cH_A\times\cH_B)^r\to(\cH_A\times\cH_B)^{r-1}$ and
$\pi':(\cH_A\times\cH_B)^r\to\cH_A\times\cH_B$ are the
projection maps sending the point $(a_1,b_1,\ldots,a_r,b_r)$
to $(a_1,b_1,\ldots,a_{r-1},b_{r-1})$ and $(a_r,b_r)$,
respectively. Consequently, we have
$J[\Phi_r]=J[\Phi_{r-1}\circ\pi]+J[\phi\circ\pi']$ and so
 \bea \label{eq:Rang}
\rank J[\Phi_r]-\rank J[\Phi_{r-1}\circ\pi]\le
\rank J[\phi\circ\pi']\le2M+2N-3.
 \eea
Thus, it suffices to prove the above claim only for the
maximal value, $r_{\rm max}$, of $r$ in the initial range.
We have done that numerically for all cases in the table by
choosing a random point $p_r$ and evaluating the rank of
$J[\Phi_r]$ at $p_r$ when $r=r_{\rm max}$. We used the same
method to prove that $\dim\cS'_l=M^2N^2-1$.
The multipartite cases (those with $n>2$) were treated similarly.

When $M=2$ and $N>1$ we have $l=2N$ and the case $r=l-1=2N-1$ is exceptional. The inequality from Theorem \ref{thm:Granica} tells
us that $\dim K\cS'_{2N-1}\le4N^2-1$. We claim that the stronger
inequality  $\dim K\cS'_{2N-1}\le4N^2-2$ is valid. Indeed, if
$\r\in K\cS'_{2N-1}$ then $L(\r^\G)=L(\r)\le2N-1$ and so we
must have $\det\r=0$ as well as $\det\r^\G=0$. Hence, the set
$K\cS'_{2N-1}$ is contained in each of the two irreducible
hypersurfaces $\det\r=0$ and $\det\r^\G=0$. Consequently, its
codimension in $H$ must be at least two. This proves our claim.
Note also that if $r=l=2N$ then the trivial inequality
$\dim K\cS'_{2N}\le4N^2$ is stronger than the one provided by
Theorem \ref{thm:Granica}. In these two cases we again use
the ranks of the Jacobian matrices to prove that these improved
bounds are attained.

The use of randomness can be avoided with additional effort.
In the bipartite case with $M=2$ and $2<N<9$ we set
$$
\begin{array}{lll}
\ket{a_k}=\ket{0}+(k-1)\ket{1},\quad &
\ket{b_k}=\ket{0}+\ket{k-1}, & k=1,\ldots,N; \\
\ket{a_k}=\ket{0}+(k-N)i\ket{1},\quad &
\ket{b_k}=\ket{0}+\ket{2N-k-1}+\ket{2N-k},
\quad & k=N+1,\ldots,2N-2; \\
\ket{a_{2N-1}}=\ket{0}+(N-1)i\ket{1},\quad &
\ket{b_{2N-1}}=i\ket{1}+\ket{N-1}; & \\
\ket{a_{2N}}=\ket{0},\quad &
\ket{b_{2N}}=\ket{N-1}, &
\end{array}
$$
where $i$ is the imaginary unit. When $N=2$ we can use the same
expressions except that we set $\ket{b_4}=\ket{0}+\ket{1}$.
Then one can verify that $J[\Phi_r]$ at the point
$p_r:=(a_1,b_1,\ldots,a_r,b_r)$ has rank $(2N+1)r$ for $r<2N-1$,
and has ranks $4N^2-2$ and $4N^2$ when $r$ is $2N-1$ and
$2N$, respectively. Hence, $\dim K\cS'_r$ must be equal to
$(2N+1)r$ for $r<2N-1$, it is $4N^2-2$ for $r=2N-1$,
and $4N^2$ for $r\ge2N$.

As another example, we consider the bipartite case $M=3$, $N=4$
in which case $r_{\rm max}=13$.
We shall select an explicit point $p_{13}$ where the rank
of the Jacobian matrix $J[\Phi_{13}]$ is equal to the upper bound
$(2M+2N-3)r_{\rm max}=11.13=143$.

Let us define the following 14 vector pairs
$(\ket{a_k},\ket{b_k})$ in $3\ox4$:
$$
\begin{array}{lcl}
\ket{a_1}=\ket{0} & \quad & \ket{b_1}=\ket{0} \\
\ket{a_2}=\ket{0}+\ket{1} & \quad & \ket{b_2}=
\ket{0}+\ket{1} \\
\ket{a_3}=\ket{0}-\ket{1} & \quad & \ket{b_3}=
\ket{0}+\ket{2} \\
\ket{a_4}=\ket{0}+i\ket{1} & \quad & \ket{b_4}=
\ket{0}-\ket{3} \\
\ket{a_5}=\ket{0}+\ket{2} & \quad & \ket{b_5}=
\ket{0}+\ket{3} \\
\ket{a_6}=\ket{0}-\ket{2} & \quad & \ket{b_6}=
\ket{0}+\ket{1}+\ket{3} \\
\ket{a_7}=\ket{0}+\ket{1}+\ket{2} & \quad & \ket{b_7}=
\ket{0}-\ket{2} \\
\ket{a_8}=\ket{0}-\ket{1}+\ket{2} & \quad & \ket{b_8}=
\ket{0}-i\ket{1} \\
\ket{a_9}=\ket{0}+(1+i)\ket{2} & \quad & \ket{b_9}=
\ket{0}+i\ket{1} \\
\ket{a_{10}}=\ket{0}+i\ket{1}-\ket{2} & \quad &
\ket{b_{10}}=\ket{0}+\ket{1}+\ket{2} \\
\ket{a_{11}}=\ket{0}+\ket{1}+i\ket{2} & \quad &
\ket{b_{11}}=\ket{0}+\ket{1}+i\ket{2} \\
\ket{a_{12}}=\ket{0}+i\ket{1}+\ket{2} & \quad &
\ket{b_{12}}=\ket{0}+i\ket{2}+\ket{3} \\
\ket{a_{13}}=\ket{0}+i\ket{1}+i\ket{2} & \quad &
\ket{b_{13}}=\ket{0}+\ket{2}+\ket{3} \\
\ket{a_{14}}=\ket{0}-i\ket{1} & \quad & \ket{b_{14}}=
\ket{0}-\ket{1}.
\end{array}
$$

Set $p_r:=(a_1,b_1,\ldots,a_r,b_r)$ for $r=1,\ldots,14$.
We have verified that the rank of $J[\Phi_{13}]$ at the point
$p_{13}$ is 143, and the rank of $J[\Phi_{14}]$ at the point
$p_{14}$ is 144. Since $\dim K\cS'_{12}=132$, each neighborhood
of the point $\Phi_{13}(p_{13})$ must contain infinitely many
separable states of length 13. A similar property is shared by
the point $\Phi_{14}(p_{14})$. However, it remains unclear
whether the state $\Phi_{13}(p_{13})$ has length 13, and the
state $\Phi_{14}(p_{14})$ the length 14.

It is tempting to conjecture that the equality
$\dim\cS'_r=\dim\cS'_{r+1}$ implies that $\cS'_r=\cS'_{r+1}$.
However, it was shown very recently that when $M=N=3$ there exist separable states of length 10. Hence, in this case
$\cS'_9\subset\cS'_{10}$ while $\dim\cS'_9=\dim\cS'_{10}=80$.
In view of this example, we shall propose a conjecture only for
bipartite systems with $M=2$ and $N$ arbitrary, in which case
$l=2N$.

 \bcj \label{conj:Granica}
Let $\cH=\cH_A\otimes\cH_B$ be the Hilbert space of a bipartite quantum system. If $\dim\cH_A=2$ and $\dim\cH_B=N$, then for any
separable state $\r$ on $\cH$ we have $L(\r)\le2N$.
 \ecj
By Theorem \ref{thm:Granica}, the conjectural bound $2N$ is the
best possible. So far, it was known that this conjecture is true
in the case of two qubits, but all other cases were open.
We have proved recently \cite{cd12qq} that it is also
true for the qubit-qutrit system.
We note that, since a separable state of rank $r\le2$ has
length $r$, we have $\cS_1=\cS'_1$ and $\cS_2=\cS'_2$.
On the other hand recall that $\cS_d=\cS$ contains an open ball
of $H'$ centered at the normalized identity matrix $(1/d)I_d$,
see \cite{zhsL98}. Consequently, $\dim\cS_d=d^2-1$.
The dimensions of the sets $\cS_r$ for $2<r<d$ are not known.
We shall compute them in the two smallest bipartite cases $2\ox2$
and $2\ox3$. The computational method from the proof below can be used to compute the dimensions of the sets $\cP_r$ consisting of all bipartite PPT states of rank at most $r$. (These are
semialgebraic sets because $\cP_r=\cD_r\cap\cD^\G$.)
Note that in the special cases that we consider here, we have $\cS_r=\cP_r$ due to the Peres-Horodecki criterion.

 \bpp \label{prop:dimS_r}
In $2\ox2$ we have $\dim\cS_3=14$. In $2\ox3$ we have
$\dim\cS_r=26,31,34$ for $r=3,4,5$, respectively.
 \epp
 \bpf
It is convenient to work with non-normalized states and so
we shall use the cones $K\cS_r$ and $K\cD_r$ instead of
$\cS_r$ and $\cD_r$.
Any non-normalized state $\r$ of rank at most $r$ can be written
as $\r=C^\dag C$, where $C$ is an $r\times2N$ matrix whose
diagonal entries are real and those below the diagonal are 0.
Let us denote by $X$ the real vector space of such matrices $C$.
Note that $\dim X=r(4N-r)$. Thus we have a surjective map
$g:X\to K\cD_r$ defined by $g(C)=C^\dag C$. We select a point
$C_0\in X$ such that the state $\r_0:=g(C_0)\in\cP_r$ and,
subject to this condition, the rank $r_0$ of the Jacobian
matrix $J_0$ of $g$ at $C_0$ is maximal.
By continuity, there is an $\epsilon>0$ such that the rank
of the Jacobian matrix is equal to $r_0$ and
$g(C)\in\cP_r$ for all $C\in X$ in the open
ball $\|C-C_0\|<\epsilon$. The image of this ball is a
submanifold of $H'$ of dimension $r_0$. Since this image is
contained in $K\cP_r=K\cS_r$, we conclude that
$\dim K\cS_r\ge r_0$.
Hence, if $r_0=r(4N-r)=\dim K\cD_r$ we can deduce that
$\dim\cS_r=\dim\cD_r$. This is indeed true in the cases
when $(N,r)$ is $(2,3)$, $(3,3)$, $(3,4)$ or $(3,5)$.
The matrices $C_0$ for these four cases can be chosen as
follows:
 \bea
\left[ \begin{array}{cccc}2&0&0&0\\0&1&1&0\\0&0&1&1
\end{array} \right],\quad [I_3~I_3],\quad
\left[ \begin{array}{cccccc}1&0&0&0&0&0\\0&1&0&1&0&0\\
0&0&1&0&0&0\\0&0&0&1&1&0
\end{array} \right],\quad
\left[ \begin{array}{cccccc}1&0&0&0&0&0\\0&1&0&1&0&0\\
0&0&1&0&0&0\\0&0&0&1&2&0\\0&0&0&0&1&1
\end{array} \right].
 \eea
 \epf

 \begin{table}
    \caption{\label{tab:dimenzije}
The dimensions of the sets $\cS'_r$ of separable states of
length at most $r$ in $d_1\ox\cdots\ox d_n$ system. The dimension increases with $r$ and reaches the maximum value $d^2-1$
for $r=l$, the bound defined by Eq. (\ref{frm:Granica}),
and remains constant afterwards.
 } \centering
 \begin{tabular}{|l|l|c|c|}
   \hline
   $d_1,\ldots,d_n$ & Initial range & Exceptional case & $d^2-1;~r\ge l$
\\\hline
   2,1 & $2;~r=1$ & & 3; $r\ge2$ \\\hline
   $2,N;~(1<N<9)$ & $(2N+1)r-1;~r<2N-1$ &
$4N^2-3;~r=2N-1$ & $4N^2-1;~r\ge2N$ \\\hline
   3,3 & $9r-1;~r<9$ & & 80; $r\ge9$ \\\hline
   3,4 & $11r-1;~r<14$ & & 143; $r\ge14$ \\\hline
   3,5 & $13r-1;~r<18$ & & 224; $r\ge18$ \\\hline
   4,4 & $13r-1;~r<20$ & & 255; $r\ge20$ \\\hline
   2,2,2 & $7r-1;~r<10$ & & 63; $r\ge10$ \\\hline
   2,2,3 & $9r-1;~r<16$ & & 143; $r\ge16$ \\\hline
   2,2,4 & $11r-1;~r<24$ & & 255; $r\ge24$ \\\hline
   2,2,2,2 & $9r-1;~r<29$ & & 255; $r\ge29$ \\\hline
 \end{tabular}
 \end{table}

We conclude this section with an interesting example.
 \bex
 \label{ex:L(rho)=MN}
{\rm Consider the separable $M\times N$ state $\s=\sum^M_{i=1}
\proj{i} \ox \r_i$ with the $\r_i>0$. Evidently, $L(\s)=\rank\s=MN$.
If $\ket{a,b}$ is a product vector and $\r=\s+\proj{a,b}$, then
$\rank\r=MN$ and we claim that $L(\r)=MN$. There is a unique $p_i>0$
such that $\s_i:=\r_i-p_i\proj{b}\ge0$ and $\rank\s_i=N-1$. Then
 \bea
 \r=\sum^M_{i=1} \proj{i}\ox\s_i + \left( \proj{a} +
\sum^M_{i=1} p_i \proj{i} \right) \ox  \proj{b}
 \eea
shows that $L(\r)\le MN$, and the claim follows. \hfill $\square$ }
 \eex

An interesting problem is to characterize the separable states $\s$,
say on $M\ox N$, such that, for every product vector $\ket{a,b}$ on
$M\ox N$, the state $\r=\s+\proj{a,b}$ satisfies the inequality
$L(\r)\le L(\s)$. The cases where $\s$ has the maximal possible
length are of course trivial. The above example shows that there
exist $\s$ with the above stated property which are not of this
trivial type. Since $L(\r)\ge\rank\r$, we infer that in
Example \ref{ex:L(rho)=MN} we always have $L(\r)=L(\s)=MN$.

We claim that there exist $\r$ and $\s$ as above such that
$L(\r)<L(\s)$. First, we show that the sum of two separable
states may have smaller length than one of the two summands.
Assume that $(M-2)(N-2)>1$. Then by Theorem \ref{thm:Granica}
we can choose a separable state $\s$ such that $L(\s)>MN$.
We can choose large $t>0$ such that $\r=tI-\s\ge0$ is separable.
Then $L(\r+\s)=MN<L(\s)$.
Second, let $k=L(\r)$ and choose a decomposition
$\r=\sum^k_{i=1}\proj{a_i,b_i}$.
Let $\s_r=\s+\proj{a_1,b_1}+\cdots+\proj{a_r,b_r}$ for
$r=0,1,\ldots,k$.
Since $L(\s_k)<L(\s)$, there exists an $r$ $(0<r\le k)$ such that
$L(\s_r)<L(\s_{r-1})$. This proves our claim.
For analytic examples of such states in $3\ox3$ see the very
recent preprint \cite{hk12dec}.

However, it is not known whether in $2\ox N$ there exists a separable state $\s$ and a state $\r=\s+\proj{a,b}$ such that
$L(\r)<L(\s)$. If such states exist, then
 \bea
L(\s)>L(\r)\ge\max(\rank\r,\rank\r^\G)\ge \max(\rank\s,\rank\s^\G).
 \eea
One can show that there exists a separable state $\s$ whose
length is larger than $\max(\rank\s,\rank\s^\G)$. Actually, analytic
examples of such $3\times3$ states can be found in \cite{hk12}. However, no such example is known in $2\ox N$. If such states do
not exist, then it would follow that Conjecture
\ref{conj:Granica} is true.

\section{Real separable states} \label{RealSepSta}

Many important states used in quantum information are {\em real}
i.e., all entries of their density matrices are real. For this
to make sense, we have to assume that we have fixed an o.n.
basis in each of the Hilbert spaces $\cH_i$, $i=1,\ldots,n$.
In this section we shall study the real separable states
on $\cH$. We have defined in the Introduction the partial
transposition operators $\G_i$ $i=1,\ldots,n$., as well as
$\G_S$ for $S\subseteq\{1,\ldots,n\}$, and the group $G$.
Recall that $H$ is the space of all Hermitian operators on
$\cH$, and $H'$ its affine subspace consisting of operators
of trace 1.

We say that an operator $\r\in H$ is {\em $G$-invariant},
and we write $\r^G=\r$, if each element of $G$ fixes $\r$.
(This is the case if and only if each $\G_i$ fixes $\r$.)
We introduce the following real spaces
 \bea
H^{\rm re} &=&  \{\r\in H:\r^*=\r\}, \\
H^G &=& \{\r\in H:\r^G=\r\}, \\
{H'}^G &=& \{\r\in H':\r^G=\r\}.
 \eea
Thus $H^{\rm re}$ is the space of all real symmetric operators on
$\cH$. It is easy to see that $H^G\subseteq H^{\rm re}$ and that
$H^G$ can be identified with the tensor product over $\bR$ of the
spaces of real symmetric operators on $\cH_i$, $i=1,\ldots,n$.
In particular, it follows that
 \bea \label{eq:dimH^G}
\dim H^G = \prod_{i=1}^n \binom{d_i+1}{2}.
 \eea
For convenience, we also set
 \bea
&& \cD^{\rm re} = \cD\cap H^{\rm re},\quad
\cS^{\rm re} = \cS\cap H^{\rm re},\quad
\cS^G = \cS \cap H^G.
 \eea
As a subset of $\cS$, the set $\cS^G$ is defined by the
equations $\r^{\G_i}=\r$, $i=1,\ldots,n$.

Let $\ket{a_i}\in\cH_i$, $i=1,\ldots,n$, be nonzero real vectors
and $\ket{\phi}=\ket{a_1,\ldots,a_n}$ the corresponding real
product vector. Then we say that $\proj{\phi}$ is a {\em real
pure product state}. These states are obviously $G$-invariant, i.e., $\proj{\phi}^{\G_S}=\proj{\phi}$ for all $S$.
We say that a state $\r$ is {\em separable over $\bR$} if it
belongs to the convex hull of the set of real pure product
states. Consequently, if a separable state $\r$ is separable over
$\bR$, then we must have $\r^G=\r$.

For instance, for the two-qubit real separable state
$\r=\proj{00}+\proj{11}+\proj{\ps}$, with
$\ket{\ps}=(\ket{01}+\ket{10})/\sqrt2$, we have $\r^\G\ne\r$. Therefore $\r$ is not separable over $\bR$.
For an explicit expression of $\r$ as the sum of four
(complex) pure product states see \cite[Eq. (137)]{App05}.

Our first result is that the above necessary condition is also
sufficient.
 \bpp \label{pp:GamaInv}
Let $\r$ be a separable state and $l=L(\r)$.
If $\r^G=\r$ then $\r$ is a sum of $2^n l$ real pure product states, and so $\r$ is separable over $\bR$.
Consequently, $\cS^G$ is the set of all states which are
separable over $\bR$.
 \epp
 \bpf
We have $\r=\sum^l_{k=1}\proj{\psi_k}$, where
$\ket{\psi_k}=\ket{a_{k1},\ldots,a_{kn}}$.
We define
\bea
\ket{b_{kj}}=(\ket{a_{kj}}+\ket{a^*_{kj}})/\sqrt{2},\quad
\ket{c_{kj}}=i(\ket{a_{kj}}-\ket{a^*_{kj}})/\sqrt{2}.
\eea
It is easy to verify that
$\proj{a_{kj}}+\proj{a^*_{kj}}=\proj{b_{kj}}+\proj{c_{kj}}$.
Since $\r^{\G_S}=\r$ for all $S\subseteq\{1,\ldots,n\}$, we have
 \bea
 2^n\r
 &=& \sum_S \sum^l_{k=1} \proj{a_{k1},\ldots,a_{kn}}^{\G_S}
 \notag\\
 &=& \sum^l_{k=1} \bigg( \proj{a_{k1}} + \proj{a_{k1}^*} \bigg)
 \ox \cdots \ox \bigg( \proj{a_{kn}} + \proj{a_{kn}^*} \bigg)
 \notag\\
 &=& \sum^l_{k=1} \bigg( \proj{b_{k1}}
  + \proj{c_{k1}} \bigg)
 \ox \cdots \ox \bigg( \proj{b_{kn}} + \proj{c_{kn}} \bigg).
 \eea
Since $\ket{b_{kj}}$ and $\ket{c_{kj}}$ are real, this completes the proof.
 \epf

For $\r\in K\cS^G$ we denote by $L^G(\r)$ the smallest integer
$k$ such that $\r$ can be written as a sum of $k$ real pure
product states. We also say that $L^G(\r)$ is the
{\em length of $\r$ over $\bR$}.
Finally, for any positive integer $r$, we set
$\cS^G_r=\{\r\in\cS^G:L^G(\r)\le r\}$.
It is immediate from the definitions that the sets
$\cS^{\rm re}$, $\cS^G$ and $\cS^G_r$ are semialgebraic.

In general, for $\r\in\cS^G$ we have $L(\r)\le L^G(\r)$.
It is an open question whether the equality always holds.
E.g., we do not know whether $\r$ in
Proposition \ref{pp:GamaInv} is a sum of $l$ real pure product states. The next theorem shows that the equality
$L(\r)=L^G(\r)$ holds in any $2\ox N$ system.

It was shown in \cite[Theorem 2]{kck00} that a $2\times N$ PPT
state $\r$ with $\r^\G=\r$ is separable. Moreover, such $\r$
admits a decomposition $\r=\sum^r_{i=1}\proj{a_i,b_i}$ with
$r=\rank\r$ and all $\ket{a_i}$ real. For real $\r$ we have
the following stronger version of this result.
 \bt
 \label{thm:2oxN,Ginvariant}
In any $2\ox N$ system, every $G$-invariant state $\r$
is separable over $\bR$. Moreover, for such $\r$
we have $L^G(\r)=\rank\r$.
 \et
 \bpf
To prove the first assertion, it suffices to consider the case
where $\r$ is a $2\times N$ state. By the theorem cited above,
$\r$ is separable. Thus, $\r\in\cS^G$ and the assertion follows
from Proposition \ref{pp:GamaInv}.

To prove the second assertion, we again may assume that $\r$
is a $2\times N$ state. It follows that $r:=\rank\r\ge N$.
Since $\r^\G=\r$, we have
 \bea
\r=\left[ \begin{array}{cc} A&B\\B&C \end{array} \right],
 \eea
where $A,B,C$ are real symmetric matrices of order $N$.
Moreover, we can assume that $C$ is invertible (see e.g.
\cite[Example 2]{cd11JPA}). By performing an invertible
real local operation on $\cH_B$, we can assume that $C=I_N$.
By performing yet another local operation on $\cH_B$, this
time with a real orthogonal matrix, we may also assume that
$B$ is a diagonal matrix, say $B=\diag(b_0,\ldots,b_{N-1})$.
Then we have
 \bea
\r=\sum_{i=0}^{N-1} \proj{\phi_i}+\proj{0}\ox A',
 \eea
where $\ket{\phi_i}=(b_i\ket{0}+\ket{1})\ox\ket{i}$ and
$A'=A-B^2$.
Since $\r\ge0$, we must have $A'\ge0$. Moreover, it is clear
that $r=N+r'$ where $r'=\rank A'$. Since $A'$ is a sum of $r'$
positive semidefinite matrices of rank one, and the
$\ket{\phi_i}$ are real product vectors, the assertion is
proved.
 \epf

The first assertion of this theorem may fail for some other
quantum systems.
To construct a $3\times3$ counter-example, we start with an UPB
$\{\ket{a_i,b_i}:i=1,\ldots,5\}$ consisting of real product
vectors. Then the state $\r=I-\sum^5_{i=1}\proj{a_i,b_i}$ is
$G$-invariant and entangled \cite{dms03}.
One can construct similarly a $2\times2\times2$ counter-example
by using the UPB from \cite[Eq. (22)]{dms03}.

We can compute the dimensions of $\cS^{\rm re}$ and $\cS^G$.
 \bpp \label{pp:Dimenzije} We have $\dim \cS^{\rm
re}=\binom{d+1}{2}-1$ and $\dim\cS^G=\prod_i \binom{d_i+1}{2}-1$.
 \epp
 \bpf Recall that there is an open ball, say $B$, in $H'$
centered at the state $\r_0:=(1/d)I_d$ such that
$B\subseteq\cS$.
Hence, the first formula follows from the facts that
$\r_0\in{H'}^{\rm re}$ and
$\dim {H'}^{\rm re}=\binom{d+1}{2}-1$, where
${H'}^{\rm re}=H'\cap H^{\rm re}$.
The second formula follows from Eq. \eqref{eq:dimH^G} by a
similar argument.
 \epf

 \begin{table}
    \caption{\label{tab:SepDim}
Lower bounds for the dimensions of the sets $\cS^G_r$
$(r=1,2,\ldots,d)$ of $G$-invariant real separable states
$\r$ with $L^G(\r)\le r$ in $M\ox N$ systems with $d=MN\le16$. }
\centering
 \begin{tabular}{|c|rrrrrrrrrrrrrrrr|}
   \hline
  ~$M,N\backslash r$~&1&2&3&4&5&6&7&8&9&10&11&12&13&14&15&16~ \\\hline
   2, 1 &~1 & 2 &&&&&&&&&&&&&& \\\hline
   2, 2 & 2 & 5 & 7 & 8 &&&&&&&&&&&& \\\hline
   2, 3 & 3 & 7 & 11 & 14 & 16 & 17 &&&&&&&&&& \\\hline
   2, 4 & 4 & 9 & 14 & 19 & 23 & 26 & 28 & 29 &&&&&&&& \\\hline
   2, 5 & 5 & 11 & 17 & 23 & 29 & 34 & 38 & 41 & 43 & 44
&&&&&& \\\hline
   2, 6 & 6 & 13 & 20 & 27 & 34 & 41 & 47 & 52 & 56 & 59
& 61 & 62 &&&& \\\hline
   2, 7 & 7 & 15 & 23 & 31 & 39 & 47 & 53 & 62 & 68 & 73
& 77 & 80 & 82 & 83 && \\\hline
   2, 8 & 8 & 17 & 26 & 35 & 44 & 53 & 62 & 71 & 79 & 86
& 92 & 97 & 101 & 104 & 106 & 107~ \\\hline
   3, 1 & 2 & 4 & 5 &&&&&&&&&&&&& \\\hline
   3, 2 & 3 & 7 & 11 & 14 & 16 & 17 &&&&&&&&&& \\\hline
   3, 3 & 4 & 9 & 14 & 19 & 24 & 29 & 32 & 34 & 35
&&&&&&& \\\hline
   3, 4 & 5 & 11 & 17 & 23 & 29 & 35 & 41 & 47 & 53 & 56
& 58 & 59 &&&& \\\hline
   3, 5 & 6 & 13 & 20 & 27 & 34 & 41 & 48 & 55 & 62 & 69
& 76 & 83 & 86 & 88 & 89 & \\\hline
   4, 1 & 3 & 6 & 8 & 9 &&&&&&&&&&&& \\\hline
   4, 2 & 4 & 9 & 14 & 19 & 23 & 26 & 28 & 29 &&&&&&&& \\\hline
   4, 3 & 5 & 11 & 17 & 23 & 29 & 35 & 41 & 47 & 53 & 56
& 58 & 59 &&&& \\\hline
   4, 4 & 6 & 13 & 20 & 27 & 34 & 41 & 48 & 55 & 62 & 69
& 76 & 83 & 90 & 96 & 98 & 99~ \\\hline
  \end{tabular}
 \end{table}

In Table II we exhibit the lower bounds for the dimensions of the
sets $\cS^G_r$ for several bipartite systems of small dimension
$(d\le16)$. In most cases we have proved that these  bounds are
equal to $\dim\cS^G_r$. To be precise, we know that the last number
(for $r=d$) of each item is correct since it is equal to the
dimension of $\cS^G$ as given by Proposition \ref{pp:Dimenzije}. The
first number (for $r=1$) is also correct, it is equal to $M+N-2$.
The numbers following it are correct as long as the difference
between the consecutive numbers is equal to $M+N-1$. This follows
from the fact that $\dim\cS^G_{r+1}-\dim\cS^G_r\le M+N-1$ for each $r$. (This inequality can be proved by a method similar
to one we used to prove \eqref{eq:Rang}.)

Let us state the general separability problem for arbitrary multipartite systems $\cH=\cH_1\ox\cdots\ox\cH_n$ and its two
special cases.

$(S_1)$ For $\r\in\cD$, decide whether $\r\in\cS$.

$(S_2)$ For $\r\in\cD\cap H^{\rm re}$, decide whether $\r\in\cS$.

$(S_3)$ For $\r\in\cD\cap H^G$, decide whether $\r\in\cS$.

Since $H^G\subseteq H^{\rm re}$, $(S_3)$ is a special case of
$(S_2)$. Next we show that $(S_1)$ can be reduced to $(S_3)$ at the expense of enlarging the dimension of the quantum system. So
the three problems in $(S_1)$, $(S_2)$ and $(S_3)$ are indeed
equivalent. Since NPT states are entangled, it suffices to
consider the PPT states only.

%
 \bpp  \label{pp:reduction-2}
Assume that each party $A_i$ consists of two parties: party
$A_{1,i}$ which has a qubit and another party $A_{2,i}$. So,
we have $\cH_i=\cH_{1,i}\ox\cH_{2,i}$.
Let $\r$ be a PPT state on $\cH_{2,1}\ox\cdots\ox\cH_{2,n}$. Then there exists a state $\s$ on the system $\cH$, which belongs to $H^G$, and is such that $\r$ is separable if and only if $\s$ is separable.
 \epp
 \bpf
The partial transposition operator $\G_i$ is the product of the
transposition operators $\G'_i$ and $\G''_i$ on $\cH_{1,i}$ and
$\cH_{2,i}$, respectively. Similarly, for any subset
$S\subseteq\{1,\ldots,n\}$, we have $\G_S=\G'_S\G''_S$. Let us
define the Hermitian operator $\s$, acting on the bipartite
composite system
$\cH=\cH_{A_{1,1},\ldots,A_{1,n}}\ox
\cH_{A_{2,1},\ldots,A_{2,n}}$,
 \bea \label{eq:sigma-2}
 \s= \sum_S \tau^{\G'_S} \ox \r^{\G''_S},
 \eea
where $\tau=\proj{a,\ldots,a}_{A_{1,1},\ldots,A_{1,n}}$,
$\ket{a}=(\ket{0}+i\ket{1})/\sqrt2$, and the summation is over all
subsets $S$ of the set $\{1,\ldots,n\}$. Since $\r$ is PPT, we have
$\s\ge0$. If $R\subseteq\{1,\ldots,n\}$, then $\G_R\G_S=\G_{R\Delta
S}$ where $R\Delta S:=(R\setminus S)\cup(S\setminus R)$ is the
symmetric difference of $R$ and $S$. Consequently, we have
 \bea \label{eq:sigma-3}
 \s^{\G_R}= \sum_S \tau^{\G'_{R\Delta S}} \ox
\r^{\G''_{R\Delta S}}=\s.
 \eea
The last equality holds because when $S$ runs through all
subsets of $\{1,\ldots,n\}$ so does ${R\Delta S}$.
Thus we have shown that $\s\in H^G$.
Since $\braket{a}{a^*}=0$, we have
$\r=\bra{a,\ldots,a}\s\ket{a,\ldots,a}$. This formula and
Eq. \eqref{eq:sigma-2} show that $\r$ is separable if and only
if $\s$ is separable.
 \epf

\section{\label{sec:conclusion}
Conclusions}

It is well known that the set of all normalized states, $\cD$,
has nonempty interior when viewed as a subset of the ambient
affine space $H'$, and so $\dim\cD=d^2-1$. No such result is known for its subsets
$\cD_r=\{\r\in\cD:\rank\r\le r\}$. First of all, we have shown that $\cD_r$ are real semialgebraic sets and so they have a
well defined dimension. Then we have given a simple formula for their dimensions (see Theorem \ref{thm:DimD_r}).

Next consider the set, $\cS$, of normalized separable states.
First we dealt with its subsets
$\cS'_r=\{\r\in\cS:L(\r)\le r\}$, where $L(\r)$ is the length
of $\r$.
We showed that each $\cS'_r$ is semialgebraic and
deduced from this fact that $\cS$ itself is semialgebraic.
We have obtained in Theorem \ref{thm:Granica} very good
upper bounds for $\dim\cS'_r$.
These bounds are not of interest when they exceed $d^2-1$.
However, in most (but not all) of the other cases that we have
computed (see Table \ref{tab:dimenzije}) these bounds are saturated.
A simple consequence of these bounds is the fact that there
exist separable states $\r$ with $L(\r)>d:=\dim\cH$.
The dimension of the subset $\cS_r=\{\r\in\cS:\rank\r\le r\}$,
is much harder to compute. We have done that for the systems
$2\ox2$ and $2\ox3$ only (see Proposition \ref{prop:dimS_r}).

We have initiated the study of real separable states. It
may be surprising that such states are not necessarily
separable over $\bR$, i.e., not necessarily
expressible as a sum of real pure product states (see the
example above Proposition \ref{pp:GamaInv}).
On the other hand we show that, among all separable states,
those which are separable over $\bR$ are characterized by the
property of being $G$-invariant.
In addition to the standard separability problem $(S_1)$,
we have formulated in section \ref{RealSepSta}
two variations $(S_2)$ and $(S_3)$ which
ask to decide whether a state $\r$ belongs to the set of
real separable states $\cS^{\rm re}$ or the set of
$G$-invariant separable states $\cS^G$, respectively.
We have shown that all three separability problems are
equivalent to each other.

Last, but not least, we have proposed a method of solving
the standard separability problem $(S_1)$ in some
low-dimensional quantum systems (see Conjectures \ref{Hip-1}
and \ref{Hip-2}). Since
these very natural conjectures are valid in the two cases
where the separability problem has been solved, namely
$2\ox2$ and $2\ox3$, we are hopeful that they may lead to
eventual analytic solution of the problem in some additional
cases. One possibility is to use the Jacobian matrix of the
map $\Phi_r$, see Eq. (\ref{eq:Phi}), where $r$ is chosen so that
$\cS'_r=\cS$. In that case, this matrix must have deficient rank
at the points mapped to the boundary of the cone $K\cS$.
This may help us to find the polynomial equations defining
the boundary of $\cS$.

The separabilty problem $(S_3)$ is the real analog of the
standard separability problem $(S_1)$. Due to the fact that
the set $\cS^G$ has much lower dimension than $\cS$, it is
very likely that in concrete cases it would be much easier to
solve $(S_3)$ than $(S_1)$. As we have shown that $(S_3)$
has a very simple solution for $2\ox N$ systems (see
Theorem \ref{thm:2oxN,Ginvariant}), the smallest open cases
are $2\ox2\ox2$ and $3\ox3$.

\acknowledgments

We thank the referee for his valuable comments.
The first author was mainly supported by MITACS and NSERC.
The CQT is funded by the Singapore MoE and the NRF as part of
the Research Centres of Excellence programme.
The second author was supported in part by an NSERC Discovery
Grant.

\end{document}